\documentclass[11pt,a4paper]{article}
\usepackage{amsfonts,amssymb,amsmath,amsthm,cite}
\usepackage{graphicx}
\usepackage{slashed}

\evensidemargin=\oddsidemargin
\title{Emission of photons by quasiparticles in Weyl semimetals}
\author{
Alexander Andrianov$^{1,2}$,  Roberto Soldati$^3$ and Dmitri Vassilevich$^{4,5}$ \\
$^1$ {\small Saint-Petersburg State University, Russia}\\
 $^2${\small Departament d'Estructura i Constituents de la Mat\`eria and }\\
{\small Institut de Ci\`encies del Cosmos (ICCUB),
Universitat de Barcelona, Spain}\\
$^3${\small   Dipartimento di Fisica e Astronomia - Universit\'{a} di Bologna}\\
{\small Istituto Nazionale di Fisica Nucleare - Sezione di Bologna, Italy}\\
 $^4${\small CMCC - Universidade Federal do ABC, SP, Brazil}\\
 $^5${\small Physics Department, Tomsk State University, Tomsk, Russia}
}
\date{}
\begin{document}
\maketitle
\begin{abstract}
We show that quasiparticles in Weyl semimetals may decay with emission of a single photon. We study the spectrum of emitted photons and estimate the decay rates.
\end{abstract}
\section{Introduction}
Many modern materials provide condensed matter realizations of the Dirac equation thus hinting to the possibility of a quantum field theory description. Among these materials, a special place belong to Weyl semimetals (see \cite{Armitage:2017cjs} for a recent review). Most of the Weyl semimetals have a constant axial vector field in the bulk, that leads to various exciting phenomena: the presence of Fermi arc states on the boundary and the chiral magnetic effect \cite{Burkov:2015hba} as a manifestation of the chiral anomaly. 

Stretching too wide the analogies to relativistic field theory may be, however, misleading. The Lorentz invariance in Weyl semiemetals is violated by the presence of the axial vector and by the difference between characteristic propagation speeds for photons and quasiparticles. Thus, the process that are strictly forbidden in a relativistic physics may become possible in Weyl semimetals. We shall study one of such processes: emission of a single photon by a quasiparticle. 

The purpose of this short note is to show that the effect exists and to estimates its magnitude. To achieve this purpose we use a lot of simplifying assumptions which include a small mass approximation and a particular initial state. As we shall see, the effect is not negligible.

This paper is organized as follows. The solutions of Dirac equation are analysed in the next section. The kinematic regions for the decay are found in section \ref{sec:kin}, while the decay probability is calculated in section \ref{sec:dec}.

\section{Spectrum of quasiparticles}
The Dirac Lagrangian that governs free propagation of quasiparticles in Weyl semimetals can be written as \cite{Armitage:2017cjs}
\begin{equation}
\mathcal{L}=\bar\psi \bigl( i\gamma^\mu \partial'_\mu -m -b_\mu \gamma^\mu \gamma^5\bigr) \psi \,.\label{Lag}
\end{equation}
Here and in what follows the prime near any vector $V$ means the rescaling of all spatial components with the Fermi velocity $v_F$, 
\begin{equation*}
V_0'=V_0,\qquad V_a'=v_F V_a,\qquad a=1,2,3. 
\end{equation*}
The axial vector $b^\mu$ is assumed to be space-like. By a suitable choice of the coordinate system it can be directed along the positive $x^3$ axis
\begin{equation}
b^\mu \equiv \delta^{\mu 3}b,\qquad b>0 .\label{bmu}
\end{equation}
The $\gamma$-matrices satisfy $\gamma^\mu\gamma^\nu + \gamma^\nu\gamma^\mu=2g^{\mu\nu}$ with $g=\mathrm{diag}\, (+1,-1,-1,-1)$.

By passing to the Fourier modes $\psi \sim e^{-ip_\mu x^\mu }$ the Dirac equation is transformed to the following form
\begin{equation}
\bigl( \gamma^\mu p'_\mu -m - b\gamma^3\gamma^5\bigr) \psi =0 \label{DirEq}
\end{equation}
To solve this equation, 
we introduce the projectors $\mathcal{P}_{\pm}=\tfrac 12 (1\pm \gamma^0\gamma^3\gamma^5)$ and corresponding
spinors $u_\pm=\mathcal{P}_\pm u$. The Dirac equation then reads
\begin{eqnarray}
&&(p_0+p_3'\gamma_5 -m\gamma_0 +b)u_+ + p'_j\gamma_0\gamma^j u_-=0\\
&&(p_0-p_3'\gamma_5 -m\gamma_0 -b)u_- + p'_j\gamma_0\gamma^j u_+=0
\end{eqnarray}
where $j=1,2$. This yields
\begin{equation}
\big( p_0^2 -(\vec{p}')^2 -m^2 -b^2 - 2b(p_3' \gamma_5 \mp m\gamma_0) \big) u_\pm =0\,.
\end{equation}
Further splitting is done with the help of the following projectors
\begin{eqnarray}
&&\mathcal{Q}_+^+=\frac 12 \left( 1 + \frac{p_3'\gamma_5-\gamma_0m}{\sqrt{(p_3')^2+m^2}} \right)\\
&&\mathcal{Q}_+^-=\frac 12 \left( 1 - \frac{p_3'\gamma_5-\gamma_0m}{\sqrt{(p_3')^2+m^2}} \right)\\
&&\mathcal{Q}_-^+=\frac 12 \left( 1 + \frac{p_3'\gamma_5+\gamma_0m}{\sqrt{(p_3')^2+m^2}} \right)\\
&&\mathcal{Q}_-^-=\frac 12 \left( 1 - \frac{p_3'\gamma_5+\gamma_0m}{\sqrt{(p_3')^2+m^2}} \right)
\end{eqnarray}
The square roots in the formulas above are all positive. 

Let us define
\begin{equation}
u_+^+=\mathcal{Q}_+^+ u_+,\quad
u_+^-=\mathcal{Q}_+^- u_+,\quad
u_-^+=\mathcal{Q}_-^+ u_-,\quad
u_-^-=\mathcal{Q}_-^- u_-
\end{equation}
Then, for $u_+^+$ and $u_-^+$ the dispersion relation reads
\begin{equation}
p_0^2 -(\vec{p}')^2-m^2-b^2 - 2b\sqrt{(p_3')^2+m^2}=0\,. \label{displus}
\end{equation}
For $u_+^-$ and $u_-^-$ we have
\begin{equation}
p_0^2 -(\vec{p}')^2-m^2-b^2 + 2b\sqrt{(p_3')^2+m^2}=0\,.\label{disminus}
\end{equation}
(See \cite{Tabert:2016} for a comprehensive analysis of dispersion relations in Weyl semimetals.)

One can easily see that $u_\pm^\pm$ are linearly independent and thus form a basis.

In this paper, we analyse the decays $e\to e\gamma$.
Let us assume that the initial and final quasiparticles obey the same dispersion law. Let us take (\ref{displus}) to be more specific. Let us denote the momentum of initial quasiparticle by $p$ and of the final - by $q$. The momentum of emitted photon is then $p-q$. We have for the $0$th components of momenta
\begin{equation}
\sqrt{{p'}_\bot^2+(b+\sqrt{{p'}_3^2+m^2})^2}-\sqrt{{q'}_\bot^2+(b+\sqrt{{q'}_3^2+m^2})^2}=|\vec{p}-\vec{q}|\label{consen}
\end{equation}
Let us use the inequality
\begin{equation}
|A-B|\geq ||A|-|B|| \label{ineq}
\end{equation}
valid for any vectors $A$ and $B$ for $A=(p'_\bot, b+\sqrt{{p'}_3^2+m^2})$ and $B=(q'_\bot,b+\sqrt{{q'}_3^2+m^2})$. 
\begin{eqnarray}
&&\sqrt{{p'}_\bot^2+(b+\sqrt{{p'}_3^2+m^2})^2}-\sqrt{{q'}_\bot^2+(b+\sqrt{{q'}_3^2+m^2})^2}\nonumber\\
&&\qquad \leq \sqrt{ (p'_\bot-q'_\bot)^2+(\sqrt{{p'}_3^2+m^2}-\sqrt{{q'}_3^2+m^2})^2}\nonumber\\
&&\qquad \leq \sqrt{ (p'_\bot-q'_\bot)^2+(p'_3-q'_3)^2}\nonumber\\
&&\qquad <|\vec{p}-\vec{q}|.\nonumber
\end{eqnarray}
To pass from the 2nd line to the 3rd, we used the same inequality applied to 2-vectors $A=(p'_3,m)$ and $B=(q'_3,m)$. The last line follows from $v_F<1$. Thus Eq.\ (\ref{consen}) cannot be satisfied. Consequently, initial and final quasiparticles have to satisfy different dispersion relations.

Let us suppose that the mass gap parameter $m$ is much smaller than the third components, $p_3'$ and $q_3'$, of rescaled momenta of the fermions involved in the decay process. In this approximation, we write
\begin{eqnarray}
&&\mathcal{Q}_+^+=\frac 12 \left( 1 + \gamma_5- \gamma_0 \frac m{p_3'} \right)\\
&&\mathcal{Q}_+^-=\frac 12 \left( 1 - \gamma_5 +\gamma_0 \frac m{p_3'} \right)\\
&&\mathcal{Q}_-^+=\frac 12 \left( 1 + \gamma_5 +\gamma_0 \frac m{p_3'} \right)\\
&&\mathcal{Q}_-^-=\frac 12 \left( 1 - \gamma_5- \gamma_0 \frac m{p_3'} \right)
\end{eqnarray}

Let us take a particular representation of the $\gamma$-matrices:
\begin{equation*}
\gamma^0=\tau_1\otimes 1_2,\qquad
\gamma^1=i\tau_2\otimes \sigma_2,\qquad
\gamma^2=-i\tau_2 \otimes \sigma_1,\qquad
\gamma^3=-i\tau_3\otimes 1_2,
\end{equation*}
where $\{ \tau \}$ and $\{ \sigma \}$ are two sets of Pauli matrices. Then
\begin{equation*}
\gamma_5=i\gamma^0\gamma^1\gamma^2\gamma^3=-\tau_2\otimes \sigma_3.
\end{equation*}

Up to normalization factors,
\begin{eqnarray}
&&u_+^+(p)= \left[ \left( \begin{array}{c} 1 \\ -i \end{array} \right) -
\frac m{2p_3'} \left( \begin{array}{c} -i \\ 1 \end{array} \right)\right] \otimes
\left( \begin{array}{c} 1\\ 0 \end{array}\right)\\
&&u_+^-(p)= \left[ \left( \begin{array}{c} 1 \\ i \end{array} \right) +
\frac m{2p_3'} \left( \begin{array}{c} i \\ 1 \end{array} \right)\right] \otimes
\left( \begin{array}{c} 1\\ 0 \end{array}\right)\\
&&u_-^+(p)= \left[ \left( \begin{array}{c} 1 \\ i \end{array} \right) +
\frac m{2p_3'} \left( \begin{array}{c} i \\ 1 \end{array} \right)\right] \otimes
\left( \begin{array}{c} 0\\ 1 \end{array}\right)\\
&&u_-^-(p)= \left[ \left( \begin{array}{c} 1 \\ -i \end{array} \right) -
\frac m{2p_3'} \left( \begin{array}{c} -i \\ 1 \end{array} \right)\right] \otimes
\left( \begin{array}{c} 0\\ 1 \end{array}\right)
\end{eqnarray}

The coupling to electromagnetic field is done by replacing $\partial_\mu \to \partial_\mu - ieA_\mu$. Thus, to compute the decay amplitudes, we have to evaluate the matrix elements $\bar u \gamma^\mu u'$ where $u$ is $u_+^+$ or $u^+_-$ and $u'$ is $u_+^-$ or $u_-^-$. Non-zero matrix elements read
\begin{eqnarray}
&\bigl( u_+^+ \bigr)^\dag (p) u_+^-(q)=\tfrac{im}{q_3'} - \tfrac{im}{p_3'}\,,\qquad &
\bigl( u_-^+ \bigr)^\dag (p) u_-^-(q)=\tfrac{im}{q_3'} - \tfrac{im}{p_3'}\\
&\bigl( u_+^+ \bigr)^\dag (p)\alpha^1 u_-^-(q)=-\tfrac{m}{q_3'} + \tfrac{m}{p_3'}\,,\qquad &
\bigl( u_-^+ \bigr)^\dag (p)\alpha^1 u_+^-(q)=\tfrac{m}{q_3'} - \tfrac{m}{p_3'}\\
&\bigl( u_+^+ \bigr)^\dag (p)\alpha^2 u_-^-(q)=\tfrac{im}{q_3'} - \tfrac{im}{p_3'}\,,\qquad &
\bigl( u_-^+ \bigr)^\dag (p)\alpha^2 u_+^-(q)=\tfrac{im}{q_3'} - \tfrac{im}{p_3'}\\
&\bigl( u_+^+ \bigr)^\dag (p)\alpha^3 u_+^-(q)=\tfrac{im}{q_3'} + \tfrac{im}{p_3'}\,,\qquad &
\bigl( u_-^+ \bigr)^\dag (p)\alpha^3 u_-^-(q)=-\tfrac{im}{q_3'} - \tfrac{im}{p_3'},
\end{eqnarray} 
where $\vec{\alpha}\equiv \gamma^0\vec{\gamma}$.
It is important to note that all matrix elements in the equations above are linear in $m$. Thus, if one is interested in the leading order of the small mass expansion only, one can compute all other quantities at $m=0$.

The modes corresponding to $u^\pm_\pm$ are not independent but rather related through the Dirac equation that has two independent solutions
\begin{eqnarray}
&& v^+(p)=(p_2'+ip_1')u_+^+ - (p_0+p_3'+b)u_-^+ \label{vp}\\
&& v^-(p)=(p_2'+ip_1')u_+^- - (p_0-p_3'+b)u_-^- \label{vm}
\end{eqnarray}
for $m=0$. 

\section{Kinematic regions for the decays}\label{sec:kin}
Let is remind that the states with dispersion relation (\ref{displus}) can decay into the states with the dispersion relation (\ref{disminus}), and vice versa. Final and initial states cannot have the same dispersion relation. It is clear that with the sign convention (\ref{bmu}) the states (\ref{displus}) allow for higher values of $p_0$ than the states (\ref{disminus}) for the same values of spatial momenta. This energy surplus is used to create a photon. Basing on these qualitative arguments (which can be confirmed by direct calculations) we conclude that the decays we are looking for is of the initial states of the type (\ref{displus}) with the spinors (\ref{vp}) to the states (\ref{disminus}) with the spinors (\ref{vm}).

Let us make a simplifying assumption that in the initial state 
\begin{equation} p_\bot=0. \label{ass}
\end{equation}
To further simplify the notations, we fix $p_3>0$. This does not affect the kinematic analysis since (\ref{displus}) and (\ref{disminus}) are not sensitive to the sign of $p_3$. We do not impose any restrictions on the momenta $q$ of the final quasiparticle. As we have explained in the previous section, in our approximation we may take $m=0$ in analysing the kinematics. The energy conservation condition yields
\begin{equation}
p_3'+b=\sqrt{{q_\bot'}^2 + \bigl( b-|q_3'| \bigr)^2}+ \sqrt{(p_3-q_3)^2 +q_\bot^2}.\label{cons1}
\end{equation} 
The momentum $q_\bot$ appears under both square roots on the right hand side of the equation above. Under the first square root, $q_\bot^2$ is multiplies by $v_F^2$ (which is a very small quantity) and thus may be neglected as compared to $q_\bot^2$ under the second square root. Thus,
\begin{equation}
p_3'+b=| b-|q_3'||+ \sqrt{(p_3-q_3)^2 +q_\bot^2}.\label{cons2}
\end{equation}
This equation can be solved for $q_\bot$ if an only if
\begin{equation}
p_3'+b-| b-|q_3'||\geq |p_3-q_3|.\label{cons3}
\end{equation}

This inequality is easy to solve. There are no solutions for $q_3<0$. For $q_3>0$ one has to distinguish two cases:
\begin{eqnarray}
&& b>q_3': \quad |p_3'-q_3'|<2p_3'v_F, \label{region1}\\
&& b<q_3': \quad  |p_3'-q_3'|<2bv_F \,. \label{region2}
\end{eqnarray}
We neglected $v_F^2$ corrections on the right had sides of both inequalities. 
Both regions are quite narrow, and the frequencies $\omega$ of emitted photons are also peaked. In the region (\ref{region1}), $\omega\simeq p_3'+q_3'\simeq 2p_3'$. While in (\ref{region2}) $\omega\simeq p_3'-q_3' +2b \simeq 2b$.

\section{Decay rates in the small mass approximation}\label{sec:dec}
We are interested in the decays where the initial fermion is in the stated described by $v^(p)$, while in the final state we have $v^-(q)$. Since we assumed that $p_\bot=0$, we can also take $u_-^+(p)$ to describe the initial state, see (\ref{vp}). Relevant matrix elements of the electromagnetic field are easily computed:
\begin{eqnarray}
&&\bigl( u_-^+ (p)\bigr)^\dag  v^-(q)=-(q_0-q_3'+b) \left(  \tfrac{im}{q_3'} - \tfrac{im}{p_3'} \right) \\
&&\bigl( u_-^+ (p)\bigr)^\dag \alpha^1  v^-(q)=(q_2'+iq_1') \left(  \tfrac{m}{q_3'} - \tfrac{m}{p_3'} \right) \\
&&\bigl( u_-^+ (p)\bigr)^\dag \alpha^2  v^-(q)=(q_2'+iq_1') \left(  \tfrac{im}{q_3'} - \tfrac{im}{p_3'} \right) \\
&&\bigl( u_-^+ (p)\bigr)^\dag \alpha^3 v^-(q)=(q_0-q_3'+b) \left(  \tfrac{im}{q_3'} + \tfrac{im}{p_3'} \right)
\end{eqnarray}
One can check, that these matrix elements satisfy the transversality condition
\begin{equation}
\bigl( u_-^+ (p)\bigr)^\dag \alpha^\mu  v^-(q) (q'-p')_\mu =0
\end{equation}
to the linear order in $m$.

Let us estimate the decay probabilities. All relevant formulas for normalizations, integration measures etc are taken from \cite{Akhiezer:1981}.
The normalized initial and final fermion states read
\begin{eqnarray}
&&\psi_{\rm i} (p) = N_{\rm i} u^+_-(p), \qquad N_{\rm i}=2^{-1/2}\label{norms1} \\
&&\psi_{\rm f} (q) = N_{\rm f} v^-(q), \qquad N_{\rm f}=(4q_0(q_0+b-q_3'))^{-1/2} \label{norms2}
\end{eqnarray}
respectively. Te emitted photon may be in two polarization states given by the formulas
\begin{eqnarray}
&& A_{(1)}(k)=N_{{\rm p},1} (0,k_2,-k_1,0), \qquad N_{\rm p}=(2k_\bot^2k_0)^{-1/2}\label{norms3} \\
&& A_{(2)}(k)=N_{{\rm p},2} (0,k_1k_3^2,k_2k_3^2,-k_3(k_1^2+k_2^2)), \qquad
N_{{\rm p},2}=(2k_0^3k_\bot^2k_3^2)^{-1/2}.
\label{norms4}
\end{eqnarray}

The differential decay probability
\begin{equation}
dw= \frac 1{(2\pi)^4} |\mathcal{A}|^2 \delta^4(p-q-k) \frac{d^3q}{(2\pi)^3}\frac{d^3k}{(2\pi)^3}\,. \label{dw}
\end{equation}
is expressed though the interaction vertex computed with normalized states
\begin{equation}
\mathcal{A}\equiv e A_i'(k) \psi_{\rm i} (p)^\dag \alpha^i  \psi_{\rm f} (q)  \,.\label{vert}
\end{equation} 

To get the full decay probability, we have to integrate (\ref{dw}) over the spatial components of $k$ and $q$. The integration over $k$ removes 3 of the 4 delta functions and enforces the spatial momentum conservation. To compute the integral over $q$, we write $d^3q = d^2q_\bot dq_3= \pi dq_\bot^2 dq_3$ (where we used the rotational symmetry of integrand to integrate over the angular variable on $q_\bot$ plane). To integrate over $dq_\bot^2$ we use the remaining delta function, so that $q_\bot^2$ has to be expressed through other momenta with the help of equation
\begin{equation}
\sqrt{(p_3-q_3)^2+q_\bot^2}+ |b-|q_3'||=p_3'+b .\label{qbot}
\end{equation}
In this equation we neglected $q_\bot'^2$ as compared to $q_\bot^2$ on the left hand side. This integration also produces a Jacobian factor $2k_0$ and enforces the integration limits for $q_3$ as prescribed by Eqs.\ (\ref{region1}) and (\ref{region2}).

The vertices (\ref{vert}) for the photons described by (\ref{norms3}) and (\ref{norms4}) read
\begin{eqnarray}
&&\mathcal{A}_{(1)}=eN_{\rm i}N_{\rm f}N_{{\rm p},1} \frac{ m( q_1^{'2}+q_2^{'2})(q_3'-p_3')}{q_3'p_3'}\\
&&\mathcal{A}_{(2)}=eN_{\rm i}N_{\rm f}N_{{\rm p},2} \frac{im}{v_F^2 p_3'q_3'}
q_\bot'^2(q_3'-p_3')\left[ (q_3'-p_3')^2 + (q_3'+p_3') (q_0-q_3'+b) \right],
\end{eqnarray}
respectively.

Since the kinematic regions (\ref{region1}) and (\ref{region2}) are very narrow, without loosing too much we may suppose that $q_3'$ and $p_3'$ are both large or both smaller than $b$. In the region (\ref{region2}) this means $p_3'>b$. For simplicity, we also assume that $p_3'-b\gg v_F b$. Here we can use the following approximate relations
\begin{eqnarray}
&&q_\bot'^2\simeq 4b^2v_F^2-(q_3'-p_3')^2 ,\\
&&q_0\simeq p_3'-b ,\\
&&q_0+b-q_3'\simeq \frac {q_\bot'^2}{2(p_3'-b)}.
\end{eqnarray}
The corrections to these formulas are of higher order in $v_F$. With these approximate formulas one can derive simple analytic formulas for the total decay probabilities. For final photons described by (\ref{norms3}), we have 
\begin{equation}
\mathcal{W}_{p_3'>b}^{(1)}=\int dw \simeq \frac {e^2 m^2 v_F^2}{8(2\pi)^9 p_3'^4} \int_{p_3-2b}^{p_3+2b} dq_3 (q_3'-p_3')^2=
\frac {2e^2 m^2 v_F^4 b^3}{3(2\pi)^9 p_3'^4} . \label{W1}
\end{equation}
Similarly, for the second photon polarization (\ref{norms4}) one obtains
\begin{equation}
\mathcal{W}_{p_3'>b}^{(2)} \simeq \frac{4 e^2 m^2 v_F^4 b^3 (3b^2-10bp_3'+15p_3'^2)}{15 (2\pi)^9 p_3'^4 (p_3'-b)^2 }
\label{W3}
\end{equation}

In the other region (\ref{region1}), when $p_3',q_3'<b$ and $b-p_3'\gg v_F b$, we can write
\begin{eqnarray}
&&q_\bot'^2\simeq 4p_3'v_F^2-(q_3'-p_3')^2 ,\\
&&q_0\simeq b-p_3' ,\\
&&q_0+b-q_3'\simeq 2(b-p_3').
\end{eqnarray}
The total decay probabilities become
\begin{equation}
\mathcal{W}_{p_3'<b}^{(1)}\simeq \frac{4 e^2 m^2v_F^6 p_3'}{15 (2\pi)^9 (b-p_3')^2}.\label{W2}
\end{equation}
for the polarization (\ref{norms3}), and
\begin{equation}
\mathcal{W}_{p_3'<b}^{(2)} \simeq \frac{16 e^2 m^2 v_F^2}{3 (2\pi)^9 p_3'}
\label{W4}
\end{equation}
for the polarization (\ref{norms4}), respectively. 

Note that these formulas have been derived assuming that $|b-p_3'|$ is finite. The apparent singularity in (\ref{W2}) and (\ref{W3}) at $b=p_3'$ signals of a crossover behaviour to a regime with a different dependence on $v_F$. 

To estimate the order of this effect, let us take $v_F=(500)^{-1}$, $m=0.1{\rm eV}$ and $p_3'=0.3{\rm eV}$. Then $\mathcal{W}_{p_3'<b}^{(2)} \simeq 800{\rm s}^{-1}$. This is a small number. However, since at least a part of positive powers of $v_F$ comes from the difference in characteristic speeds of the fermions and of the photons, just taking into account the refraction index of the material could significantly improve the result. Also, getting rid of the small mass approximation is going to increase the decay probability. One may hope to get in this way the lifetime of the order of about tens microseconds. To compare, we note that this is already of the same order as the characteristic time scale of electronic cooling through interaction with phonons in Weyl and Dirac semimetals \cite{Lundgren:2015}. This makes the effect phenomenologically significant, especially taking into account a very specific spectrum of emitted photons that have their frequencies sharply peaked at $2p_3'$. 

The other decay probabilities  (\ref{W1}), (\ref{W2}) and (\ref{W3}) are damped by higher powers of $v_F^2$ and thus are less important. 

We have to stress, that we have studied just a single possible relative orientation (parallel) of the initial state momentum and the axial vector $b$.
 
\section{Conclusions}
The main message of this work is that in contrast to the intuition obtained through working in Lorentz invariant field theories, the quasiparticles in Weyl semimetals may decay emitting a single photon. We have studied this effect in a small mass approximations and demonstrated that it is small but not too small. We have argued that by giving up the small mass approximation and by taking into account the refraction index of the bulk of Weyl semimetals may lead to a considerable enhancement of the decay probability. Besides, it is interesting to study the effects of chemical potential and of the temperature. 

\subsection*{Acknowledgments}
This work was started during the visit of A.A. to UFABC supported by the S\~ao Paulo Research Foundation (FAPESP), project 2014/23772-1. The work of A.A. was supported by the project RFBR 18-02-00264. The work of D.V. was supported in parts by FAPESP, project 2016/03319-6, by CNPq, projects 428951/2018-0 and 305594/2019-2, by the RFBR project 18-02-00149-a and by the Tomsk State University Competitiveness Improvement Program.

\end{document}